\documentclass[12pt]{article}

\usepackage{amsmath,amsfonts,amssymb,cite,graphicx,bbm}

\def\bea{\begin{eqnarray}}
\def\eea{\end{eqnarray}}
\def\be{\begin{equation}}
\def\ee{\end{equation}}
\def\ba{\begin{array}}
\def\ea{\end{array}}

\newcommand{\ib}{{\bar{i}}}
\newcommand{\jb}{{\bar{j}}}

\newcommand{\mV}{\mathcal V}

\newcommand{\I}{\mathrm{i}}
\DeclareMathOperator{\Det}{\Delta}

\newcommand{\C}{{\mathbbm{C}}}

\numberwithin{equation}{section}

\usepackage[colorlinks]{hyperref}

\begin{document}

\setlength\arraycolsep{2pt}

\renewcommand{\theequation}{\arabic{section}.\arabic{equation}}
\setcounter{page}{1}

\setlength\arraycolsep{2pt}

\begin{titlepage}

\begin{center}

\vskip 0.4 cm

{\LARGE  \bf Constraints from de Sitter metastability in heterotic string compactifications}

\vskip 0.7cm

{\large 
Dirk Rathlev
}

\vskip 0.5cm

{\it
Institut f\"ur Theoretische Physik, Universit\"at Z\"urich, \\
 CH-8057 Z\"urich, Switzerland\\
}

\vskip 0.8cm

\end{center}

\begin{abstract}

We study the possibility of obtaining metastable de Sitter vacua of heterotic string theory compactified on a Calabi-Yau threefold which are classical and simple in the K\"ahler moduli sector of the theory. For this, we exploit a known necessary condition on the K\"ahler potential in $\mathcal{N}=1$-supergravity, which we, under the assumption that only moduli fields contribute to supersymmetry breaking, express in terms of a tensorial eigenvalue problem for the Calabi-Yau triple intersection tensor. For three-dimensional moduli spaces we are able to identify the discriminant of the Calabi-Yau intersection tensor in the analysis, generalizing a known result for two-dimensional moduli spaces. We also discuss explicit examples and possible generalizations.

\end{abstract}

\end{titlepage}

\newpage

\noindent
 \pagenumbering{arabic} 

\section{Introduction}
A major challenge of realistic model building in string theory is the construction of theories with supersymmetry breaking metastable de Sitter vacua in their low-energy effective supergravity descriptions. Such models are desirable as they provide a simple explanation for the accelerated expansion of the universe convincingly suggested by cosmological measurements. Unfortunately, de Sitter vacua are notoriously difficult to obtain in a string theoretic setting. It has been made clear by the formulation of several no-go theorems (see for example \cite{hep-th/0007018,1003.0029,0711.2512,1107.2925,1110.0545}) why naive attempts of obtaining de Sitter models necessarily have to fail in many cases.\par\medskip

A natural possibility is to generate a tiny positive cosmological constant via various types of small corrections to a leading-order Minkowski vacuum  \cite{Becker:2002nn,Balasubramanian:2004uy,Parameswaran:2006jh,Palti:2008mg,Berg:2007wt,1204.0807}. However, these attempts are often facing difficulties controlling higher-order contributions. A promising alternative approach is to start with an anti de Sitter vacuum and then `uplift' this vacuum to a de Sitter vacuum by adding additional hard supersymmetry breaking contributions \cite{Kachru:2003aw,Silverstein:2007ac,0901.0683}. These additional contributions however often complicate the supergravity description of models obtained in this way.\par\medskip

Despite significant progress in overcoming the difficulties of both the approaches sketched above (see for example \cite{1203.1750,1208.3208}), it still appears useful -- from a model building as well as from a conceptual point of view -- to study situations where de Sitter vacua arise naturally at the leading order in the low-energy effective supergravity description. This has for example been done for a broad range of type II string models in Ref.\ \cite{0812.3551,0812.3886,Haque:2008jz,0907.2041,1103.4858,1112.3338}. A more general program to identify and study the main obstacles for the appearance of metastable de Sitter vacua has been started in \cite{GRS1}. It has been found that the critical parameter is the average mass of the sGoldstinos, which is not allowed to become negative. This result directly excludes leading order de Sitter vacua in situations where only the dilaton or a single K\"ahler modulus contributes to supersymmetry breaking, confirming the earlier result in \cite{0402088}. The more general case where the moduli space can be factorized into one-dimensional submanifolds and thus the K\"ahler potential is given by
\begin{align}
  \label{eq:separable_potential}
  K = -\sum_i n_i\log\left(T^i+\overline{T}^i\right),\qquad \sum_i n_i=3
\end{align}
is also excluded by the same constraint. This analysis has been extended to two contributing K\"ahler moduli, spanning a non-trivial two-dimensional scalar manifold, in the case of compactifications of heterotic string theory as well as orientifold compactifications of type IIb string theory in \cite{Covi:2008ea,CGGPS}. The result of the analysis is an explicit topological constraint on the Calabi-Yau compactification manifold, which has to be fulfilled for metastable de Sitter vacua to exist. The main goal of this paper is to determine how much of this analysis can be carried over to the more general case where $p>2$ K\"ahler moduli contribute to supersymmetry breaking. We only consider the case of heterotic compactifications. Most of the corresponding results for orientifold compactifications of type IIb string theory can be easily obtained by exploiting the duality between these two theories, which in this case simply amounts to a sign change of the relevant quantity $\omega$ defined in Eq.\ \eqref{eq:omega_def} (see for example the appendix of \cite{Gross-2009-PhdThesis}).\par\medskip

This paper is organized as follows. In Section \ref{sec:metastability} we briefly recall the constraint on the K\"ahler potential in metastable de Sitter  $\mathcal{N}=1$-supergravity coming from the Goldstino multiplet. In the low-energy effective supergravity theory obtained from heterotic string theory compactified on a Calabi-Yau threefold, this constraint can be encoded in the sign of a homogeneous function called $\omega$. Section \ref{sect:twoandthree} studies the properties of $\omega$ and reformulates the task of determining its sign as a tensorial eigenvalue problem. This formulation is exploited to extend the known result for two-dimensional moduli spaces to the three-dimensional situation. We then apply the results of Section \ref{sect:twoandthree} to the study of three-dimensional examples in Section \ref{sec:examples}. We conclude in Section \ref{sec:conclusion}.

\section{Metastability in supergravity}
\label{sec:metastability}
We start by briefly reviewing the strategy developed in \cite{GRS1,GomezReino:2007qi,GomezReino:2008bi,Gross-2009-PhdThesis,Covi:2008ea,CGGPS} to study the existence of metastable vacua with a non-negative cosmological constant in $\mathcal{N}=1$-supergravity. We assume that vector multiplets do not play a significant role in supersymmetry breaking and therefore only consider the chiral multiplets.\par\medskip

The Lagrangian of $n$ chiral multiplets is completely specified by a single real function $G$ of the superfields $\Phi^i$ and their conjugates $\bar{\Phi}^\ib$. This function can be decomposed as $G=K+\ln\left|W\right|^2$, with the real K\"ahler potential $K$ and the holomorphic superpotential $W$. The scalar fields $\phi^i$ and $\bar{\phi}^\ib$ span a K\"ahler manifold with metric
\begin{align}
  g_{i\bar{j}}=K_{i\bar{j}}=\partial_i\partial_{\bar{j}}K=\frac{\partial^2 K}{\partial \phi^i\partial\bar{\phi}^{\bar{j}}}.
\end{align}
The scalar potential is given by
\begin{align}
	\label{eq:sc_potential}
	V = e^G\left(G^iG_i-3\right),
\end{align}
where $G_i = \partial_i G$. The vacuum condition then reads
\begin{align}
	\label{eq:vac_cond}
	e^G\left(G_i+G^k\nabla_iG_k\right)+G_i V = 0,
\end{align}
where $\nabla_i$ denotes the K\"ahler-covariant derivative. The corresponding mass squared matrix is given by the vacuum expectation value of the Hessian of the scalar potential:
\begin{align}
  \label{eq:massmatrix}
	M^2 = \begin{pmatrix}
			V_{i\jb} & V_{ij}\\
			V_{\ib\jb} & V_{\ib j}
	      \end{pmatrix}.
\end{align}
Here and in the following we omit the $\langle\dots\rangle$-brackets for quantities evaluated at the vacuum. Using Eq.\ \eqref{eq:sc_potential} and Eq.\ \eqref{eq:vac_cond}, the entries of $M^2$ can be worked out and are given by (see \cite{Covi:2008ea})
\begin{align}
	\label{eq:Vijbar}
	V_{i\bar{j}} &= e^G\left(G_{i\bar{j}}+\nabla_iG_k\nabla_{\bar{j}}G^k-R_{i\bar{j}m\bar{n}}G^mG^{\bar{n}}\right)+\left(G_{i\bar{j}}-G_iG_{\bar{j}}\right)V\\
	V_{ij} &= e^G\left(2\nabla_iG_j+G^k\nabla_i\nabla_jG_k\right)+\left(\nabla_iG_j-G_iG_j\right)V,
\end{align}
where $R_{i\bar{j}m\bar{n}}$ is the Riemann tensor of the K\"ahler geometry. For the vacuum to be metastable, the mass matrix in Eq.\ \eqref{eq:massmatrix} has to be positive definite. A necessary condition for this to be the case is that the upper-left block $V_{i\jb}$ is positive definite. We now consider the projection of this block onto the direction of the Goldstino $G^i=g^{\jb i}\partial_\jb G$ in the space of chiral fermions to isolate the contribution of the Goldstino multiplet to the mass matrix:
\begin{align}
	\lambda := e^{-G}V_{i\jb}G^iG^\jb.
\end{align}
This is a positive combination of eigenvalues of $V_{i\jb}$ and should therefore be positive if $M^2$ is a positive definite matrix. Using Eq.\ \eqref{eq:Vijbar} and Eq.\ \eqref{eq:vac_cond} to calculate $\lambda$ more explicitly, one finds \cite{GRS1}
\begin{align}
	\lambda = 2g_{i\jb}G^iG^\jb-R_{i\jb m\overline{n}}G^iG^\jb G^mG^{\overline{n}}.
\end{align}
The superpotential in string compactifications is usually quite complicated due to all sorts of non-perturbative contributions and classical background effects. Thus, we will think of the superpotential as generic, which can in principle be tuned to a suitable value. The vacuum expectation value of $G^i$ does depend on the superpotential and hence can be varied by varying $W$. The coefficients in $\lambda$ however depend only on the K\"ahler geometry and thus the condition
\begin{align}
	\max_{G^i}\left\{\lambda\right\} = \max_{G^i}\left\{2g_{i\jb}G^iG^\jb-R_{i\jb m\overline{n}}G^iG^\jb G^mG^{\overline{n}}\right\}\overset{!}{>}0
\end{align}
for the existence of a supersymmetry breaking metastable vacuum does only depend on the K\"ahler potential $K$ and gives a constraint on possible K\"ahler potentials in viable theories.\par\medskip

The cosmological constant in supergravity theories is given by the vacuum expectation value $V$ of the scalar potential. To incorporate the requirement of a non-negative cosmological constant, it is useful to rewrite $\lambda$ as (see again \cite{Covi:2008ea})
\begin{align}
	\lambda &= -\frac{2}{3}e^{-G}V\left(e^{-G}V+3\right)+\sigma\\
	\label{eq:def_sigma}
	\text{with}\qquad\sigma &= \left[\frac{2}{3}g_{i\jb}g_{m\overline{n}}-R_{i\jb m\overline{n}}\right]G^iG^\jb G^mG^{\overline{n}}.
\end{align}
The sign of $\sigma$ does only depend on the direction of $G^i$, not on its length. Assume there is a vector $G^i$ such that $\sigma(G^i)>0$. By rescaling $G^i\to rG^i$ one can always achieve $V(rG^i)=0$ (i.e.\ a Minkowski vacuum) at which $\sigma(rG^i)>0$ and thus $\lambda(rG^i)$ is still positive. By rescaling $r$ a bit further, one gets $V(rG^i)>0$ and -- if the change in $r$ is small enough -- $\lambda(rG^i)>0$ still holds, proving that the condition for the existence of a metastable de Sitter vacuum is satisfied. Conversely, if $\sigma<0$ for all directions of $G^i$, $\lambda$ can never be made positive as long as $V(G^i)>0$ holds. In summary:
\begin{align}
	V>0 \text{ and } \lambda>0 \text{ is possible} \Leftrightarrow \sigma>0\text{ is possible}.
\end{align}
Hence, a necessary condition for the existence of metastable de Sitter vacua is completely encoded in the sign of $\sigma$.\par\medskip

As an example, we briefly sketch the argument given in Ref.\ \cite{GRS1} to exclude classical metastable de Sitter vacua for the K\"ahler potential in Eq.\ \eqref{eq:separable_potential}. The Riemann tensor in this case is completely diagonal and its entries are given by
\begin{align}
  R_{i\ib i\ib} = g_{i\ib}g_{i\ib}R_i,
\end{align}
where $R_i$ is the scalar curvature of the $i$-th one-dimensional submanifold, given by
\begin{align}
  R_i = \frac{K_{ii\ib\ib}}{K_{i\ib}^2}-\frac{K_{ii\ib}K_{i\ib\ib}}{K_{i\ib}^3} = \frac{2}{n_i}.
\end{align}
Parameterizing the Goldstino vector $G^i$ such that $G_iG^i=\Theta_i^2$, where $\Theta_i$ are real numbers satisfying $\sum_i\Theta_i^2=1$, we obtain
\begin{align}
  \sigma = \frac{2}{3}-\sum_i R_i \Theta^4.
\end{align}
Extremizing $\sigma$ under the constraint $\sum_i\Theta_i^2=1$ gives for its maximum
\begin{align}
  \sigma_{max} = \frac{2}{3}-\frac{1}{\sum_i R_i^{-1}} = \frac{2}{3}-\frac{2}{\sum_i n_i},
\end{align}
which vanishes in the no-scale case $\sum_i n_i=3$, thus proving that metastable de Sitter vacua do not exist.

\subsection{Compactifications of heterotic string theory}
\label{sec:hetcompacts}
We will study compactifications of heterotic string theory on Calabi-Yau manifolds in the following. The low-energy limits of these theories are known to give $\mathcal{N}=1$-supergravity models \cite{1985NuPhB.25846C} and we can therefore use the constraint described above to restrict their admissible K\"ahler potentials. We will assume that only moduli fields participate significantly in supersymmetry breaking. This assumption has attracted some phenomenological interest recently, in particular in type IIb models \cite{aci,aciII}. Nonetheless it should eventually be dropped to make the analysis more universal. Work in that direction has been done in \cite{FSscal} by the addition of matter fields to the analysis. It has been found that the study of metastability can (under some assumptions) be decomposed into two `orthogonal' parts and one of these only involves the moduli fields. Thus, the assumption in this paper is justified in the sense that the results are directly applicable to a subset of the generalized problem.\par\medskip

As another simplification we will also neglect bundle moduli and their interplay with the metric moduli. Under this assumption, the moduli space parameterizing the deformations of the Calabi-Yau manifold $Y_6$ consists of the deformations of the complex structure and the deformations of the K\"ahler form. Locally, the moduli space then factorizes as
\begin{align}
	\mathcal{M}_{CY} = \mathcal{M}^{ks}\times\mathcal{M}^{cs},
\end{align}
where the first factor consists of the K\"ahler structure deformations and the second one of the complex structure deformations. Remarkably, it turns out that both $\mathcal{M}^{ks}$ and $\mathcal{M}^{cs}$ are itself K\"ahler manifolds, not only when they are combined to give $\mathcal{M}_{CY}$ \cite{CdlO}.\par\medskip

We will assume that only K\"ahler moduli contribute to supersymmetry breaking. Note that, as complex structure moduli and K\"ahler moduli are interchanged by mirror-symmetry (see for example \cite{Greene199015}), the analysis for the case where only complex structure moduli take part in supersymmetry breaking would be identical.\par\medskip

The classical volume of the Calabi-Yau manifold $Y_6$ is denoted by $\mathcal{V}$. It holds \cite{CdlO}
\begin{align}
	\mathcal{V} = \frac{4}{3} \int_{Y_6} J\wedge J\wedge J,
\end{align}
where $J$ is the K\"ahler $(1,1)$-form. $J$ is a harmonic form and can therefore be written as $J=v^i w_i$, where $w_i$, $i=1,\dots,\,h^{1,1}$ is a basis of the $H^{1,1}$-cohomology group of $Y_6$. In string theory compactifications an additional geometric structure arises, a real two-form $B=b^i w_i$, which is connected to the metric by supersymmetry. As argued in \cite{CdlO}, natural local coordinates on the moduli space $\mathcal{M}^{ks}$ are given by $T^i=v^i+\I b^i$ and the classical volume can be written as
\begin{align}
	\label{eq:volume_het}
	\mathcal{V} = \frac{1}{6}d_{ijk}\left(T^i+\overline{T}^i\right)\left(T^j+\overline{T}^j\right)\left(T^k+\overline{T}^k\right).
\end{align}
The symmetric rank-3 tensor $d_{ijk}$ is defined by
\begin{align}
	d_{ijk} := \int_{Y_6} w_i\wedge w_j\wedge w_k,
\end{align}
and consists of the (real) Calabi-Yau triple intersection numbers.\par\medskip

In the large-volume limit, i.e.\ if the volume of the Calabi-Yau is large compared to the string scale, the K\"ahler potential of $\mathcal{M}^{ks}$ is simply given by
\begin{align}
	K = -\log \mathcal{V}.
\end{align}
Note that in particular the dimension of the moduli space $\mathcal{M}^{ks}$ (which we will call $p$ throughout this article) is given by the $(1,1)$-Betti number of the Calabi-Yau manifold $Y_6$:
\begin{align}
	p:=\dim \mathcal{M}^{ks} = \dim H^{1,1}(Y_6).
\end{align}
The computation of the K\"ahler metric and the Riemann tensor can be found in \cite{Covi:2008ea} and the result reads
\begin{align}
	\label{eq:metric_heter}
	g_{ij} &= -\frac{\mathcal{V}_{ij}}{\mathcal{V}}+\frac{\mathcal{V}_i\mathcal{V}_j}{\mathcal{V}^2} = e^K d_{ijk}K^k+K_iK_j\\
	R_{ijmn} &= g_{ij}g_{mn}+g_{in}g_{mj}-e^{2K}d_{imp}g^{pq}d_{qjn}
\end{align}
with
\begin{align}
	\label{eq:K_up}
	K^i &= -\left(T^i+\overline{T}^i\right)\\
	\label{eq:K_down}
	K_i &= -\frac{1}{2}e^K d_{ijk}\left(T^j+\overline{T}^j\right)\left(T^k+\overline{T}^k\right).
\end{align}
Note that we dropped the bar above indices referring to derivatives with respect to a complex-conjugated quantity. As the K\"ahler potential only depends on real fields, every derivative can be thought of as a derivative w.r.t.\ a real quantity.\par\medskip

As can be directly checked using e.g.\ Eq.\ \eqref{eq:K_up}, Eq.\ \eqref{eq:K_down} and Eq.\ \eqref{eq:volume_het}, the supergravity theory satisfies the \textit{no-scale property}:
\begin{align}
 \label{eq:noscale}
  K_iK^i=3.
\end{align}
In models satisfying the no-scale property it always holds that $\sigma(K^i)=0$. It turns out to be useful to exploit the specialty of the $K^i$-direction by explicitly decomposing the Goldstino direction $G^i$ into a part parallel to $K^i$ and a part $N^i$ orthogonal to $K^i$:
\begin{align}
	\label{eq:Goldstino_decomp}
	G^i = \alpha K^i+N^i.
\end{align}
Using the no-scale property Eq.\ \eqref{eq:noscale}, the projector onto the orthogonal complement of $K^i$ is given by
\begin{align}
  \label{eq:projector_with_g}
  P^{ij} = g^{ij}-\frac{1}{3}K^iK^j.
\end{align}
The function $\sigma$ defined in Eq.\ \eqref{eq:def_sigma} can be decomposed into a negative-semidefinite part and a part which only involves the orthogonal direction $N^i$ (in fact this also works in a much broader class of models, see for example \cite{Covi:2008ea} for the general discussion):
\begin{align}
	\sigma = -2s_is^i + \omega,
\end{align}
where $s^i$ and $\omega$ are given by
\begin{align}
	s^i &= \alpha\overline{N}^i+\overline{\alpha}N^i-\frac{1}{2}e^KP^{ij}d_{jmn}N^m\overline{N}^n\\
	\label{eq:omega_def}
	\omega &= \left(-\frac{4}{3}g_{ij}g_{mn}+\frac{1}{3}g_{im}g_{jn}+\frac{1}{2}e^{2K}d_{ijp}P^{pq}d_{qmn}+e^{2K}d_{imp}P^{pq}d_{qjn}\right)N^i\overline{N}^jN^m\overline{N}^n.
\end{align}
The term $-2s_is^i$ is always non-positive and thus a necessary condition for the positivity of $\sigma$ is the positivity of $\omega$. The computation and analysis of $\omega$ is the main goal of this study. It can be shown (see \cite{Rathlev:masterthesis}) that the positivity of $\omega$ and the positivity of $\sigma$ are actually equivalent conditions in this case, at least for $p=2$- and $p=3$-dimensional moduli spaces.\par\medskip

Physically acceptable points $K^i=-\left(T^i+\overline{T}^i\right)$ on the moduli space have to satisfy three conditions:
\begin{align}
  \label{eq:cond1}
 \mathcal{V}(K^i)&>0,\\
 \label{eq:cond2}
 g(K^i)&>0,\\
 \label{eq:cond3}
 \max_{N^i,\,N^iK_i=0}\omega(K^i,N^i)&>0,
\end{align}

where the first condition states the positivity of the volume of the Calabi-Yau threefold $Y_6$, the second ensures the positivity of the kinetic energy of the moduli fields (which is also connected to the positivity of the spacetime metric of the compactified dimensions) and the third condition is the metastability condition. Note that Eq.\ \eqref{eq:cond2} and Eq.\ \eqref{eq:cond3} are invariant under $K^i\to-K^i$ while Eq.\ \eqref{eq:cond1} changes its sign. Thus, to determine the physically acceptable regions of the moduli space, it is sufficient to only solve Eq.\ \eqref{eq:cond2} and Eq.\ \eqref{eq:cond3} and then swap the overall orientation of $K^i$ if necessary to also solve Eq.\ \eqref{eq:cond1}. Note also that one eigenvalue of $g_{ij}$ is always positive due to Eq.\ \eqref{eq:noscale} and hence $g_{ij}$ is positive definite if and only if $g_{ij}n_\alpha^i\overline{n}_\alpha^j>0$ for an orthogonal basis $n_\alpha^i,\,\alpha=1,\dots,\,p-1$, of the orthogonal complement of $K^i$. This gives $p-1$ constraints and together with Eq.\ \eqref{eq:cond3} there is a total of $p$ conditions to be satisfied. We will in the following assume that the points satisfying $g>0$ have already been identified and only study the additional constraint coming from $\omega>0$.

\section{Metastability analysis of heterotic compactifications}
\label{sect:twoandthree}
To check if $\omega$ can be positive for a suitable Goldstino direction $G^i=\alpha K^i+N^i$, its global maximum as a function of the orthogonal direction $N^i$ has to be determined. Since $\omega$ is a homogeneous function of $N^i$, we can assume that $N^i$ is normalized: $N^i\overline{N}_i=1$.\par\medskip

To proceed, we fix an arbitrary real orthonormal basis of the subspace orthogonal to $K^i$, i.e.\ a set of $p-1$ vectors $n_\alpha^i$ satisfying
\begin{align}
	\label{eq:real_orth_basis}
	K_in_\alpha^i=0,\quad n_{\alpha i}n_\beta^i = \delta_{\alpha\beta},\quad \bar{n}_\alpha^i=n_\alpha^i \quad\text{ for }\, \alpha,\,\beta=1,\dots,\,p-1.
\end{align}
In terms of these basis vectors, the projector $P^{ij}$ onto the orthogonal complement of $K^i$ can be written as
\begin{align}
	\label{eq:projector_p}
	P^{ij} = \sum_{\alpha=1}^{p-1}n_\alpha^in_\alpha^j.
\end{align}
A general unit vector $N^i$ orthogonal to $K^i$ can be parameterized as
\begin{align}
	\label{eq:param_N_p}
	N^i = \sum_{\alpha=1}^{p-1} e^{\I\varphi_\alpha}c_\alpha n_\alpha^i
\end{align}
with real phases $\varphi_\alpha$ and real $c_\alpha$ satisfying
\begin{align}
	\sum_{\alpha=1}^{p-1} c_\alpha^2=1.
\end{align}

With Eq.\ \eqref{eq:projector_p} and Eq.\ \eqref{eq:param_N_p} $\omega$ in Eq.\ \eqref{eq:omega_def} can be written as
\begin{align}
	\omega &= -\frac{4}{3}+\frac{1}{3}\left|\sum_\alpha c_\alpha^2e^{2\I\varphi_\alpha}\right|^2+\frac{1}{2}\sum_\alpha\left(\sum_{\beta\gamma}c_\beta c_\gamma D_{\alpha\beta\gamma}  e^{\I(\varphi_\beta-\varphi_\gamma)}\right)^2\nonumber\\
	&\qquad\qquad\qquad\qquad\qquad\quad+\sum_\alpha\left|\sum_{\beta\gamma}c_\beta c_\gamma D_{\alpha\beta\gamma}  e^{\I(\varphi_\beta+\varphi_\gamma)}\right|^2\nonumber\\
	&= -\frac{4}{3}+\frac{1}{3}\sum_{\alpha\beta}c_\alpha^2c_\beta^2\cos(2\varphi_{\alpha\beta})\nonumber\\
	&\quad\qquad+\sum_{\alpha\beta\gamma\delta\eta}c_\beta c_\gamma c_\delta c_\eta D_{\alpha\beta\gamma}D_{\alpha\delta\eta}\left[\frac{1}{2}\cos(\varphi_{\beta\delta}-\varphi_{\gamma\eta})+\cos(\varphi_{\beta\delta}+\varphi_{\gamma\eta})\right],
	\label{eq:omega_heter_p_gen}
\end{align}
where we defined the symmetric rank 3 tensor
\begin{align}
  \label{eq:transverse_intersect}
  D_{\alpha\beta\gamma}:=e^Kd_{ijk}n_\alpha^in_\beta^jn_\gamma^k
\end{align}
and used the abbreviation $\varphi_{\beta\delta}:=\varphi_\beta-\varphi_\delta$.\par\medskip

We will in the following assume that $N^i$ is real, i.e.\ that all complex phases $\varphi_\alpha$ vanish. This assumption is fully justified for $p<4$: For $p=2$, only a global phase is present in $\omega$, which drops out immediately. For $p=3$, it can be verified (see appendix \ref{sect:complex_phase}) that it is safe to set the complex phases $\varphi_\alpha$ to zero in the sense that it does not spoil the validity of the positivity analysis in this case. With vanishing complex phases, $\omega$ simplifies to
\begin{align}
	\label{eq:omega_gen_wophases}
	\omega &= -1+\frac{3}{2}\sum_{\alpha\beta\gamma\delta\eta}c_\beta c_\gamma c_\delta c_\eta D_{\alpha\beta\gamma}D_{\alpha\delta\eta}
		= -1+\frac{3}{2}\sum_{\alpha=1}^{p-1} D_{\alpha NN}^2.
\end{align}
Here and in the following we use an extension of the abbreviation in Eq.\ \eqref{eq:transverse_intersect}:
\begin{align}
	D_{uvw}:=e^Kd_{ijk}u^iv^jw^k
\end{align}
for vectors $u^i$, $v^i$ and $w^i$, while a subscript $\alpha$ on the left-hand side stands for a contraction with $n_\alpha^i$.\par\medskip

The vector $N^i$ in Eq.\ \eqref{eq:param_N_p} can also be parameterized by $p-2$ angles $\vartheta_\beta$ (e.g.\ using spherical coordinates). $N^i\equiv N^i(\vartheta_\beta)$ can be supplemented with $p-2$ additional unit vectors $N_2^i,\dots,\, N_{p-1}^i$ such that $\{N^i,\,N_\alpha^i\}$ forms an orthonormal frame orthogonal to $K^i$. The projector $P^{ij}$ in Eq.\ \eqref{eq:projector_p} can then be expressed as
\begin{align}
  P^{ij} = N^iN^j+\sum_{\alpha=2}^{p-1} N_\alpha^iN_\alpha^j \equiv \sum_{\alpha=1}^{p-1} N_\alpha^iN_\alpha^j,
\end{align}
where we defined $N_1^i\equiv N^i$. This amounts to the substitution $n_\alpha^i\to N_\alpha^i$ in Eq.\ \eqref{eq:omega_gen_wophases}, i.e.
\begin{align}
	\omega &= -1+\frac{3}{2}\sum_{\alpha=1}^{p-1} D_{N_\alpha NN}^2.
\end{align}
The derivative of $N_\alpha^i$ w.r.t.\ one of the $p-2$ angles parameterizing $N^i$ (and thus all $N_\alpha^i$) is again orthogonal to $K^i$ (because $K^i$ does not depend on any of these angles) and we can write
\begin{align}
	 \label{eq:coeff_matrix_omegacrit}
        \partial_\alpha N^i \equiv \frac{\partial}{\partial\vartheta_\alpha} N^i &= \sum_{\gamma=2}^{p-1} a_{\alpha\gamma} N_\gamma^i\\
        \label{eq:Nbetapartialalpha}
        \partial_\alpha N_\beta^i \equiv \frac{\partial}{\partial\vartheta_\alpha} N_\beta^i&= -a_{\alpha\beta} N^i +\sum_{\gamma=2}^{p-1}a_{\alpha\beta\gamma}N_\gamma^i \quad\text{for}\quad \beta\geq2,
\end{align}
with some matrix $a_{\alpha\gamma}$ and tensor $a_{\alpha\beta\gamma}$ of coefficient functions whose precise form depend on the parameterization of $N^i$. The tensor $a_{\alpha\beta\gamma}$ is antisymmetric in its last two indices (as can be seen via integration by parts):
\begin{align}
        \label{eq:antisymmetry_a}
        a_{\alpha\beta\gamma} &= -a_{\alpha\gamma\beta}.
\end{align}
Using Eq.\ \eqref{eq:coeff_matrix_omegacrit}, Eq.\ \eqref{eq:Nbetapartialalpha} and Eq.\ \eqref{eq:antisymmetry_a}, a short calculation shows that a critical point of $\omega$ has to satisfy
\begin{align}
        \label{eq:omega_crit_mat}
        \sum_{\beta=2}^{p-1} D_{NNN_\beta}\left[\delta_{\alpha\beta}D_{NNN}+D_{NN_\alpha N_\beta}\right] = 0,\quad\alpha=2,\dots,\,p-1.
\end{align}
Consequently, a subset of the critical points of $\omega$ is given by the solutions of
\begin{align}
        \label{eq:omega_crit_p}
        D_{NNN_\beta}=0,\quad\beta=2,\dots,\,p-1,
\end{align}
i.e.\ by the critical points of $D_{NNN}$ (cf.\ Eq.\ \eqref{eq:coeff_matrix_omegacrit}). We denote these critical points by $\vec{\vartheta}_j$, $j=1,\dots,q_c$ and proceed by assuming that the global maximum of $\omega$ is indeed contained in this set of critical points. This assumption can be explicitly verified in the three-dimensional case (see Sect.\ \ref{sec:fourth_crit} in the appendix). At the critical points given by Eq.\ \eqref{eq:omega_crit_p}, $\omega$ reads
\begin{align}
  \omega = -1+\frac{3}{2}D_{NNN}^2.
\end{align}

\subsection{Tensorial eigenvalue formulation}
\label{sec:tens_ev_prob}
We will show that solving Eq.\ \eqref{eq:omega_crit_p} is equivalent to solving the tensorial eigenvalue problem
\begin{align}
	\label{eq:ev_problem}
  e^Kd_{ijk}v^jv^k = \lambda  I_{ijk}v^jv^k,\qquad i=1,\dots,p,
\end{align}
where
\begin{align}
  \label{eq:def_I_p}
  I_{ijk} = \frac{1}{3}\left(K_ig_{jk}+K_{j}g_{ki}+K_kg_{ij}\right).
\end{align}
A nonzero vector $v^i$ is called a \textit{tensorial eigenvector} of $e^Kd_{ijk}$ if there exists a $\lambda \in \C$ (the corresponding \textit{tensorial eigenvalue}) such that $v^i$ solves Eq.\ \eqref{eq:ev_problem}. The role of the determinant in linear (matrix) eigenvalue problems is now played by the discriminant (which is therefore sometimes called the \textit{hyperdeterminant}\footnote{This hyperdeterminant should however not be confused with the \textit{Cayley hyperdeterminant}, which usually differs from the discriminant by factors of other invariants.}) $\Det\left[d_{ijk}\right]$, defined as the minimal-degree homogeneous polynomial in the tensor components $d_{ijk}$ which satisfies
\begin{align}
	\label{eq:def_discr_nev}
	\Det\left[d_{ijk}\right]=0\Leftrightarrow \exists v^i\neq0:\, d_{ijk}v^jv^k=0.
\end{align}
These properties fix the discriminant uniquely up to a normalization. The normalization is typically chosen such that
\begin{align}
	\label{eq:hypdet_norm}
	\Det\left[E_{ijk}\right]=1,
\end{align}
where $E_{ijk}=\delta_{ij}\delta_{jk}$ is the unit tensor. Closed form expressions for the discriminant are known for small $p$ (see e.g.\ Eq.\ \eqref{eq:discrim} for $p=2$ and Ref.\ \cite{aronhold} for $p=3$). Formulas to systematically compute discriminants for larger $p$ can be found in Ref.\ \cite{gelfand1994discriminants}, but the resulting expressions are extremely lengthy.\par\medskip

The discriminant is an invariant of $d_{ijk}$, meaning that under the transformation $d_{ijk}\to d_{ijk}' = d_{lmn}U^l_{~i}U^m_{~j}U^n_{~k}$ with some $p\times p$-matrix $U$ the discriminant transforms as
\begin{align}
  \Det\left[d_{ijk}\right] \to \Det\left[d_{ijk}'\right] = \left(\det U\right)^{\frac{3}{p}\deg \Det\left[d_{ijk}\right]} \Det\left[d_{ijk}\right],
\end{align}
where
\begin{align}
	\label{eq:degree_hypdet}
	\deg \Det\left[d_{ijk}\right]=p\cdot 2^{p-1}
\end{align}
denotes the degree of the discriminant in the tensor entries $d_{ijk}$. Equation \eqref{eq:def_discr_nev} implies that all tensorial eigenvalues are roots of the \textit{characteristic polynomial}.
\begin{align}
	\label{eq:char_polynom}
	\Det\left[e^Kd_{ijk}-\lambda I_{ijk}\right]=0.
\end{align}
Together with Eq.\ \eqref{eq:hypdet_norm} this implies that up to normalization the discriminant is equal to the product of all tensorial eigenvalues:
\begin{align}
	\label{eq:hypdet_prod_ev}
	\Det\left[I_{ijk}\right]\cdot\prod_{n=1}^{p\cdot 2^{p-1}} \lambda_n= \Det\left[e^Kd_{ijk}\right],
\end{align}
where $\lambda=\lambda_n$ are the solutions of Eq.\ \eqref{eq:ev_problem}.\par\medskip

We now prove the following statement: If $N^i$ satisfies
\begin{align}
        \label{eq:DNNN_crit_p2}
  D_{NNN}=e^K d_{ijk}N^iN^jN^k \text{ extremal},\quad N_iK^i=0,\quad N_iN^i=1
\end{align}
then three solutions of the eigenvalue problem
\begin{align}
    \label{eq:ev_prob_p}
    e^Kd_{ijk}v^jv^k = \lambda I_{ijk}v^jv^k,
\end{align}
with the right-hand side defined in Eq.\ \eqref{eq:def_I_p}, are given by
\begin{align}
        \label{eq:v_decomp_p}
        v^i&=\alpha K^i+N^i,\\
  \label{eq:alpha_p}
  \text{where}\quad \alpha^2 &= \frac{1-\lambda}{6+9\lambda}\\
  \label{eq:D_NNN_alpha}
  D_{NNN} &= 2\alpha(\lambda-1)\\
  \label{eq:DNNN_lambda_p}
  \Rightarrow D_{NNN}^2 &= 4\frac{\left(1-\lambda\right)^3}{6+9\lambda}.
\end{align}
Conversely, solutions of the eigenvalue problem of the form $v^i=\alpha K^i+N^i$ give critical points of $D_{NNN}$ satisfying Eq.\ \eqref{eq:D_NNN_alpha}.\par\medskip

Eq.\ \eqref{eq:alpha_p} and Eq.\ \eqref{eq:D_NNN_alpha} are obtained by plugging the ansatz in Eq.\ \eqref{eq:v_decomp_p} into the eigenvector equations and multiplying these with $K^i$ and $N^i$ respectively. To prove that $v^i$ actually solves the eigenvector problem, we introduce a Lagrange multiplier $\mu$ to implement the constraint $N_iK^i=0$ and consider the function
\begin{align}
  f(N^i,\mu) = \frac{D_{NNN}-\mu I_{ijk}N^iN^jN^k}{\left(N_iN^i\right)^{3/2}}.
\end{align}
Extremizing this function is equivalent to extremizing $D_{NNN}$ under the constraints in Eq.\ \eqref{eq:DNNN_crit_p2}.\par\medskip

Differentiating with respect to $N^i$ gives
\begin{align}
  0=\frac{\partial}{\partial N^l}f &= 3\frac{N_iN^ie^Kd_{ljk}N^jN^k-N_l D_{NNN}-N_iN^i\mu I_{ljk}N^jN^k+\mu N_l I_{ijk}N^iN^jN^k}{\left(N_iN^i\right)^{5/2}}\nonumber\\
  &= 3\left[e^Kd_{ljk}N^jN^k-N_lD_{NNN}- I_{ljk}N^jN^k\right],
  \label{eq:DNNN_crit_lagrange}
\end{align}
where in the second line we used both constraints from Eq.\ \eqref{eq:DNNN_crit_p2} and $\mu=1$, which follows from multiplying the first line with $K^l$.\par\medskip

Multiplication of $e^Kd_{ijk}v^jv^k$ with an arbitrary vector $n^i$ orthogonal to $K^i$ and use of Eq.\ \eqref{eq:DNNN_crit_lagrange}, \eqref{eq:v_decomp_p} and \eqref{eq:D_NNN_alpha} gives
\begin{align}
        \label{eq:proof_ev_prob}
  e^Kd_{ijk}n^iv^jv^k&\overset{\eqref{eq:v_decomp_p}}{=}2\alpha n^iN_i + e^K d_{ijk}n^iN^jN^k\nonumber\\
  &\overset{\eqref{eq:DNNN_crit_lagrange}}{=} n^iN_i \left(2\alpha+D_{NNN}\right) \overset{\eqref{eq:D_NNN_alpha}}{=} n^iN_i2\alpha\lambda \overset{\eqref{eq:v_decomp_p}}{=} \lambda I_{ijk}n^iv^jv^k,
\end{align}
which proves the claim.\par\medskip

The converse statement is proven by deducing Eq.\ \eqref{eq:DNNN_crit_lagrange} from Eq.\ \eqref{eq:proof_ev_prob}.

\subsection{$p=2$-dimensional moduli spaces}
\label{sec:2_dim}
The two-dimensional case $p=2$ has been studied in Ref.\ \cite{Covi:2008ea}. If $p=2$, the subspace orthogonal to $K^i$ is 1-dimensional and the (up to orientation) only real unit vector orthogonal to $K^i$ is
\begin{align}
  \label{eq:N_fixed}
	N^i = \frac{1}{\sqrt{3\det g}}\begin{pmatrix}K_2\\-K_1\end{pmatrix}.
\end{align}
The orthogonal part $N^i$ contains no free parameters and no extremization needs to be performed. $\omega$ can be obtained directly by a somewhat tedious calculation. The result is
\begin{align}
	\label{eq:omega_twofields}
  \omega = \frac{9}{8}e^{4K}\frac{\Det\left[d_{ijk}\right]}{\det g^3},
\end{align}
where $\Det\left[d_{ijk}\right]$ is a homogeneous polynomial in the intersection numbers defined by the formula
\begin{align}
	\label{eq:discrim}
  \Det\left[d_{ijk}\right] := d_{111}^2d_{222}^2 + 4d_{111}d_{122}^3 +4d_{222}d_{112}^3 - 3d_{112}^2d_{122}^2 -6d_{111}d_{222}d_{112}d_{122}.
\end{align}
This polynomial is the discriminant of the homogeneous polynomial $f(x^1,x^2)=d_{ijk}x^ix^jx^k$. Note that the usual definition of the discriminant of a binary cubic includes an additional factor of $-27$.\par\medskip

In physical regions of the moduli space, $\det g>0$ has to hold. In these regions, $\omega$ is positive if and only if the discriminant $\Det\left[d_{ijk}\right]$ is positive.\par\medskip

The same result can be found by using the tensorial eigenvalue problem in Eq.\ \eqref{eq:ev_prob_p}. Note that in the $p=2$-dimensional case, $N^i$ in Eq.\ \eqref{eq:v_decomp_p} is fixed by Eq.\ \eqref{eq:N_fixed} and together with $K^i$ it spans the whole vector space. Thus, Eq.\ \eqref{eq:v_decomp_p}, Eq.\ \eqref{eq:alpha_p} and Eq.\ \eqref{eq:D_NNN_alpha} completely specify all tensorial eigenvectors with non-vanishing orthogonal component. The eigenvalues are given by the solutions of Eq.\ \eqref{eq:DNNN_lambda_p} (where now $D_{NNN}\equiv D_{111}$) and their product satisfies
\begin{align}
  \prod_{i=2}^4 \lambda_i = 1-\frac{3}{2}D_{111}^2 = -\omega.
\end{align}
There is one additional eigenvalue
\begin{align}
  \lambda_1=-\frac{2}{3}
\end{align}
with an eigenvector in $K^i$-direction:
\begin{align}
	e^K d_{ijk}K^jK^k = -2 K_i = -\frac{2}{3} I_{ijk}K^jK^k,
\end{align}
where we used Eq.\ \eqref{eq:K_down} and the definition of $I_{ijk}$ in Eq.\ \eqref{eq:def_I_p}. Using Eq.\ \eqref{eq:hypdet_prod_ev}, we find
\begin{align}
	\label{eq:omega_twofields_temp}
	-\frac{2}{3}\left(-\omega\right) = \prod_{i=1}^4 \lambda_i = \frac{\Det\left[e^Kd_{ijk}\right]}{\Det\left[I_{ijk}\right]}.
\end{align}
Using Eq.\ \eqref{eq:discrim} to calculate the discriminant of $I_{ijk}$ gives
\begin{align}
  \Det\left[I_{ijk}\right] = \frac{4}{3}\det g^3,
\end{align}
which back in Eq.\ \eqref{eq:omega_twofields_temp} reproduces the result in Eq.\ \eqref{eq:omega_twofields}.

\subsection{$p=3$-dimensional moduli spaces}
\label{sec:3_dim}
We now consider the three-dimensional situation. To simplify the notation, we explicitly parameterize $N^i$ by
\begin{align}
  \label{eq:param_N_3}
  N^i = \cos\vartheta\,n_1^i+\sin\vartheta\,n_2^i
\end{align}
with real orthonormal vectors $n_1^i$ and $n_2^i$ as in Eq.\ \eqref{eq:real_orth_basis}. The real unit vector $N_2^i$ orthogonal to $K^i$ and $N^i$ is then given by
\begin{align}
  M^i := N_2^i = -\sin\vartheta\,n_1^i+\cos\vartheta\,n_2^i = \frac{\partial}{\partial\vartheta}N^i.
\end{align}
According to Eq.\ \eqref{eq:omega_crit_mat}, a local extremum of $\omega$ has to satisfy
\begin{align}
	\label{eq:om_fact_2}
  D_{NNM}(D_{NNN}+D_{NMM}) = 0,
\end{align}
i.e.\ either
\begin{align}
	\label{eq:first_poss}
  D_{NNM} = 0
\end{align}
or
\begin{align}
	\label{eq:second_poss}
  D_{NNN}+D_{NMM} = 0
\end{align}
holds. It can be explicitly checked (the argument can be found in Section \ref{sec:fourth_crit} in the appendix) that the second possibility, which is in fact a linear equation in $\tan\vartheta$, can be discarded in the positivity analysis, as the corresponding critical point is not the global maximum of $\omega$. Hence, we are left with the task of determining the solutions of
\begin{align}
  \label{eq:crit_points_3}
	D_{NNM}(\vartheta) = \frac{1}{3}\frac{\partial}{\partial\vartheta} D_{NNN}= 0,
\end{align}
i.e.\ finding the critical points of $D_{NNN}(\vartheta)$. This equation is cubic in $\tan\vartheta$. Its solutions can be determined analytically, though the explicit results unfortunately are not very illuminating. The tensorial eigenvalue formulation derived in Section \ref{sec:tens_ev_prob} however turns out to be more suitable for studying the properties of the critical points of $\omega$.

\subsubsection{A product formula}
\label{sec:product_formula}
We now exploit the formulation of the maximization problem for $\omega$ in terms of the tensorial eigenvalue problem Eq.\ \eqref{eq:ev_prob_p} to obtain the generalization of Eq.\ \eqref{eq:omega_twofields} in the $p=3$-dimensional case. For this, we explicitly identify all tensorial eigenvalues specified by Eq.\ \eqref{eq:ev_problem}.\par\medskip

As in the two-dimensional case, one eigenvalue is always given by 
\begin{align}
  \label{eq:first_ev_3}
  \lambda_1=-\frac{2}{3}
\end{align}
with an eigenvector in $K^i$-direction:
\begin{align}
	e^K d_{ijk}K^jK^k = -2 K_i = -\frac{2}{3} I_{ijk}K^jK^k.
\end{align}
The other eigenvalues are given by the solutions of
\begin{align}
  \label{eq:lambda}
  D_{NNN}^2 = \frac{4}{3}\frac{(1-\lambda)^3}{3\lambda+2},
\end{align}
where $D_{NNN}$ is evaluated at one of its three critical points. For every critical point of $D_{NNN}$, this equation gives three tensorial eigenvalues. If the critical point is real, exactly two of them will be complex and the third will be real. In addition, we have $D_{NNN}^2>\frac{2}{3}$ and thus $\omega>0$ if and only if the real eigenvalue is negative. The product of the three eigenvalues from the $j$-th critical point $\lambda_{3j+1}, \lambda_{3j+2}, \lambda_{3j+3}$ satisfies
\begin{align}
  \prod_{i=1}^3 \lambda_{3j+i} = 1-\frac{3}{2}D_{NNN}^2(\vartheta_j) = -\omega(\vartheta_j)\quad\text{for}\quad j=1,\,2,\,3.
\end{align}
The three critical points of $D_{NNN}$ give $9$ eigenvalues via Eq.\ \eqref{eq:lambda} and together with the one in Eq.\ \eqref{eq:first_ev_3} we found $10$ eigenvalues. Thus, the characteristic polynomial in Eq.\ \eqref{eq:char_polynom} is of degree $10$. However, the discriminant is of degree $12$ in the tensor components and we must conclude that the right-hand side $I_{ijk}$ is (doubly) degenerated:
\begin{align}
	\Det \left[I_{ijk}\right]=0.
\end{align}
This spoils the validity of Eq.\ \eqref{eq:hypdet_prod_ev}, which we otherwise could have used to derive a generalization of Eq.\ \eqref{eq:omega_twofields}. To repair this flaw, the eigenvalue problem Eq.\ \eqref{eq:ev_problem} has to be regularized, i.e.\ we have to replace the tensor on the right-hand side:
\begin{align}
  \label{eq:regularization}
  I_{ijk}\to I_{ijk}^\varepsilon = I_{ijk}+\varepsilon\, \delta I_{ijk}
\end{align}
such that for $\varepsilon>0$
\begin{align}
  \Det\left[I_{ijk}^\varepsilon\right] \not= 0.
\end{align}
One possibility is
\begin{align}
  \delta I_{ijk} = E_{ijk} = \delta_{ij}\delta_{jk}.
\end{align}
This substitution deforms the eigenvalues found above only by terms of order $\varepsilon$ but introduces two additional eigenvalues. These can be calculated by an expansion in $\varepsilon$ for the eigenvector $v^i$ and the eigenvalue $\lambda$ of the form
\begin{align}
	\label{eq:perturbed_v_3}
	v^i &= \alpha K^i + \beta n_1^i + \gamma n_2^i,\\
	\label{eq:perturbed_ev_3}
	\alpha&=\varepsilon\alpha_1+\mathcal{O}(\varepsilon^2),\quad \beta=\pm\I\gamma+\varepsilon\beta_1+\mathcal{O}(\varepsilon^2),\quad \lambda = \frac{\mu}{\varepsilon}+\mathcal{O}(\varepsilon^0).
\end{align}
Plugging Eq.\ \eqref{eq:perturbed_v_3} into the eigenvector equations gives for the product of the two new eigenvalues
\begin{align}
	\lambda_{2}\lambda_{3}=|\lambda_{2}|^2 = \frac{(D_{111}-3D_{122})^2+(D_{222}-3D_{112})^2}{\varepsilon^2|\I C_1+C_2|^2}+\mathcal{O}(\varepsilon^{-1}),
\end{align}
where
\begin{align}
  |\I C_1+C_2|^2 = \left|\sum_{j=1}^3\left(\I n_1^j+n_2^j\right)^3\right|^2.
\end{align}
Putting everything together and using Eq.\ \eqref{eq:hypdet_prod_ev} we finally find
\begin{align}
	\prod_{j=1}^3 \omega(\vartheta_j) &= -\frac{\prod_{i=4}^{12}\lambda_i}{\lambda_1\lambda_2\lambda_3}\nonumber\\
	&=  \frac{3}{2}e^{12K}\Det\left[d_{ijk}\right]\frac{|\I C_1+C_2|^2}{(D_{111}-3D_{122})^2+(D_{222}-3D_{112})^2}\left[\lim_{\varepsilon\to 0}\frac{\varepsilon^2}{\Det\left[I_{ijk}^\varepsilon\right]}\right],
\end{align}
where $\vartheta_j$ are the three critical points of $D_{NNN}$.\par\medskip

It remains to compute the quantity $\lim_{\varepsilon\to 0} \left[\varepsilon^{-2}\Det\left[I_{ijk}^\varepsilon\right]\right]$. This can be done either via a brute-force approach using the explicit expressions for the discriminant derived in \cite{aronhold} or via a perturbative expansion for the eigenvectors of $I_{ijk}^\varepsilon$. In either way, one finds for the leading order result in $\varepsilon$
\begin{align}
  \Det\left[I_{ijk}^\varepsilon\right] = \varepsilon^2\frac{4^3}{3^4}|\I C_1+C_2|^2 \det g^6+\mathcal{O}(\varepsilon^3),
\end{align}
which gives
\begin{align}
	\label{eq:omega_prod}
	\prod_{j=1}^3 \omega(\vartheta_j) &= \frac{243}{128}\,\frac{e^{12K}}{\det g^6}\,\frac{\Det\left[d_{ijk}\right]}{(D_{111}-3D_{122})^2+(D_{222}-3D_{112})^2}.
\end{align}
Equation \eqref{eq:omega_prod} is the three-dimensional generalization of the two-dimensional result Eq.\ \eqref{eq:omega_twofields}. Unfortunately, it has less predictive power as it is only a statement about the \textit{product} of a subset of the critical points of $\omega$, though we know that this subset contains the global maximum. However, one important conclusion can be drawn, namely
\begin{align}
  \label{eq:pos_discr}
  \Det\left[d_{ijk}\right] > 0 \quad\Rightarrow\quad \omega = -1+\frac{3}{2}\left(D_{NNN}^2+D_{NNM}^2\right) > 0
\end{align}
at at least one critical point of $\omega$, providing that $g>0$ is satisfied. Note however that the converse does not necessarily hold: for negative $\Det\left[d_{ijk}\right]$ either one or all critical points of $\omega$ on the left-hand side of Eq.\ \eqref{eq:omega_prod} can be negative and the number of negative critical points may even vary on the moduli space.\par\medskip

To verify Eq.\ \eqref{eq:omega_prod} and to study the possible existence of a converse of Eq.\ \eqref{eq:pos_discr} we performed a numerical study. Note that the discriminant $\Det\left[d_{ijk}\right]$ can be written as (see \cite{aronhold})
\begin{align}
  \Det\left[d_{ijk}\right] = T^2\left[d_{ijk}\right] - S^3\left[d_{ijk}\right],
\end{align}
where $S\left[d_{ijk}\right]$ and $T\left[d_{ijk}\right]$ denote the \textit{Aronhold invariants}\footnote{In the literature, $S$ is sometimes defined with an additional factor of $\frac{1}{4}$.}, well-known invariants of cubic polynomials in three dimensions of degree 4 and 6 in the tensor components respectively (see \cite{duistermaat2010discrete} for a modern exposition).\par\medskip

\begin{figure}
 \includegraphics[scale=1.3]{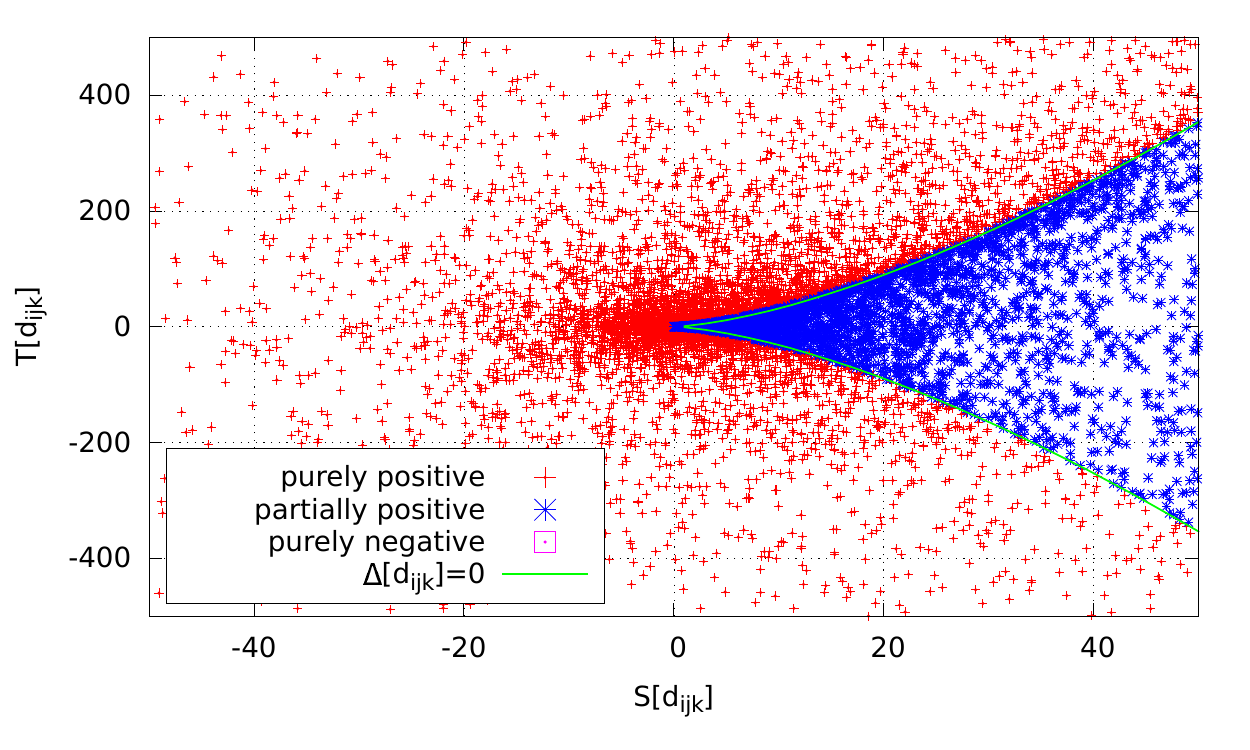}
\caption{$S-T$-plane showing the metastability classification of $10^4$ randomly generated heterotic string models. Note that there are no purely negative points.}
\label{fig:heterotic}
\end{figure}

A numerical code has been used to randomly generate sets of intersection numbers and to classify the resulting models as either \textit{purely positive}, meaning that every physical point on the moduli space fulfills $\max_{\vartheta}\omega(\vartheta) > 0$ or as \textit{partially positive}, meaning that only a subset of the physical points allow for a positive $\omega$. In the generic case, $S\neq0,\,T\neq0,\,\Det\neq0$, no model has been found which would qualify for a \textit{purely negative} classification, i.e.\ which does not allow metastable de Sitter vacua at all. Plotting all generated models in the $S-T$-plane (see Fig.\ \ref{fig:heterotic}) shows the correctness of Eq.\ \eqref{eq:pos_discr} and demonstrates that its converse does not hold: all generated models with $\Det\left[d_{ijk}\right]<0$ allow metastable de Sitter vacua at a proper subset of the moduli configurations satisfying the basic requirement $g>0$.

\subsection{Higher-dimensional moduli spaces}
\label{sec:general_case}
The result in the last section can in principle be generalized to arbitrary $p$; the explicit calculations however become quite involved and have not yet been carried out completely. In this section we briefly sketch the steps required for a generalization of Eq.\ \eqref{eq:omega_prod} and anticipate the final result.\par\medskip

In the three-dimensional case, we found three families of tensorial eigenvalues: One eigenvalue $\lambda_1=-\frac{2}{3}$ corresponding to an eigenvector in $K^i$ direction, nine eigenvalues corresponding to critical points of $\omega$ via Eq.\ \eqref{eq:DNNN_lambda_p} and two eigenvalues introduced by the regularization in Eq.\ \eqref{eq:regularization}. The same families exist in the general case: We have one  eigenvalue $\lambda_1=-\frac{2}{3}$ with an eigenvector in $K^i$ direction, $3q_c=3\,(2^{p-1}-1)$ eigenvalues corresponding to the critical points of $\omega$ in Eq.\ \eqref{eq:omega_crit_p} via Eq.\ \eqref{eq:DNNN_lambda_p} and $q_r=(p-3)2^{p-1}+2$ eigenvalues introduced by the regularization of the right-hand side. In total the number of eigenvalues is equal to the degree of the discriminant $p\cdot 2^{p-1}$, see Eq.\ \eqref{eq:degree_hypdet}. In complete analogy to the three-dimensional case one can derive a product formula for the values of $\omega$ evaluated at a subset of its critical points. It reads
\begin{align}
  \label{eq:omega_prod_p}
  \prod_{j=1}^{q_c} \omega(\vec{\vartheta}_j) = A_p\,\Det \left[e^Kd_{ijk}\right],
\end{align}
where $\vec{\vartheta}_j$ denotes the solutions of Eq.\ \eqref{eq:omega_crit_p} and $A_p$ is a function which has not been calculated yet. It is defined by
\begin{align}
  A_p = \frac{2}{3}\left[\lim_{\varepsilon\to 0}\frac{\varepsilon^{q_r}}{\Det\left[I_{ijk}^\varepsilon\right]}\right]\prod_{j=2+3q_c}^{p\cdot 2^{p-1}}\mu_j^{-1},
\end{align}
where the product in the last factor contains all eigenvalues introduced by the regularization via $\lambda_j=\varepsilon \mu_j$ (cf.\ Eq.\ \eqref{eq:perturbed_ev_3}).\par\medskip

Ref.\ \cite{Rathlev:masterthesis} contains a partial argument why $A_p>0$ should hold. Pending the completion of that argument, we can again draw the conclusion that
\begin{align}
  \label{eq:main_corrolary_p}
  \Det\left[d_{ijk}\right] > 0 \quad\Rightarrow\quad \max_{\vartheta}\omega > 0
\end{align}
holds in every physical region of the moduli space.

\section{Explicit examples}
\label{sec:examples}
Using the machinery developed above, we now study three important classes of examples in more detail.

\subsection{Perturbations around zero eigenvalues}
The eigenvalue problem Eq.\ \eqref{eq:ev_problem} is in general difficult to solve explicitly. Often feasible however is the determination of zero eigenvectors, i.e.\ solving the problem for $\lambda=0$, which exist if and only if the discriminant of $e^K d_{ijk}$ vanishes. In this case, $\omega$ vanishes at the corresponding critical point and it may be of interest to take subleading contributions to the intersection tensor into account.\par\medskip

Let $v_0^i$ denote a zero eigenvector:
\begin{align}
  e^K d_{ijk}v_0^jv_0^k = 0.
\end{align}
We now determine the solutions of the perturbed problem in which
\begin{align}
  \label{eq:pertubed_dijk}
  d_{ijk}\to d_{ijk}+\varepsilon c_{ijk}
\end{align}
to leading order in the small parameter $\varepsilon$. Plugging the ansatz
\begin{align}
  v^i = v_0^i+\varepsilon v_1^i,\qquad \lambda=\varepsilon\mu
\end{align}
into the eigenvector equations \eqref{eq:ev_problem} and multiplying with $v_0^i$ we find\footnote{For simplicity, we assume that the zero eigenspace (or rather zero eigenvariety) is locally one-dimensional. If this is not the case, the zero eigenvector $v_0^i$ is fixed by the condition that the linear system of equations specifying $v_1^i$ has a solution.}
\begin{align}
  \label{eq:lambda_order_1}
  \mu = e^K \frac{c_{ijk}v_0^iv_0^jv_0^k}{I_{ijk}v_0^iv_0^jv_0^k}.
\end{align}
Expanding Eq.\ \eqref{eq:lambda} in $\varepsilon$, we obtain
\begin{align}
  D_{NNN}^2 = \frac{2}{3}\left(1-\frac{9}{2}\varepsilon\mu\right)+\mathcal{O}(\varepsilon^2)
\end{align}
and thus for the corresponding $\omega$
\begin{align}
  \label{eq:omega_perturbed}
  \omega = -1+\frac{3}{2}D_{NNN}^2 = -\frac{9}{2}\varepsilon\mu+\mathcal{O}(\varepsilon^2).
\end{align}
As an example, we consider $p=3$ and the simple factorizing volume
\begin{align}
	\mathcal{V} = -d_{123}K^1K^2K^3.
\end{align}
This model has three vanishing eigenvalues with eigenvectors
\begin{align}
	w_1^i=\begin{pmatrix}1\\0\\0\end{pmatrix},\quad w_2^i=\begin{pmatrix}0\\1\\0\end{pmatrix},\quad w_3^i=\begin{pmatrix}0\\0\\1\end{pmatrix}
\end{align}
and $\omega$ therefore vanishes at the corresponding critical points:
\begin{align}
	\omega(\vartheta_1)=\omega(\vartheta_2)=\omega(\vartheta_3)=0.
\end{align}
If we perturb the model as in Eq.\ \eqref{eq:pertubed_dijk}, we can use Eq.\ \eqref{eq:lambda_order_1} and Eq.\ \eqref{eq:omega_perturbed} to obtain
\begin{align}
  \label{eq:pertubation_example}
	\omega(\vartheta_1)=-e^K\frac{9}{2}\varepsilon\frac{c_{111}}{K_1g_{11}},\quad\omega(\vartheta_2)=-e^K\frac{9}{2}\varepsilon\frac{c_{222}}{K_2g_{22}},\quad\omega(\vartheta_3)=-e^K\frac{9}{2}\varepsilon\frac{c_{333}}{K_3g_{33}}.
\end{align}

\subsection{Diagonal intersection numbers for $p=3$}
\label{sec:ex_diag_3}
The next case we are going to study are $p=3$-dimensional models with purely diagonal intersection numbers, i.e.\ $d_{iii}\not=0$ and all other $d_{ijk}$ vanish.\par\medskip

In this case $\mathcal{V}_{ij}=\partial_i\partial_j \mathcal{V}$ is diagonal and it holds
\begin{align}
	\det g &= -\frac{1}{2}e^{3K}d_{111}d_{222}d_{333}K^1K^2K^3.
\end{align}
We choose orthonormal basis vectors orthogonal to $K^i$ by
\begin{align}
	\label{eq:n1_ex_3}
	n_1^i = \frac{1}{\sqrt{C_{12}}}\begin{pmatrix}K_2\\-K_1\\0\end{pmatrix}
\end{align}
and
\begin{align}
	n_2^i = \sqrt{\frac{\det g}{3}}\varepsilon^{ijk}K_jn_{1k},
\end{align}
where $C_{12}$ is a normalization constant given by
\begin{align}
	\label{eq:n_norm}
	C_{12} &= \frac{1}{4}e^{3K}d_{111}d_{222}K^1K^2\left(d_{111}(K^1)^3+d_{222}(K^2)^3\right).
\end{align}
With these choices we have $D_{122}=0$. If $N^i$ is parameterized as in Eq.\ \eqref{eq:param_N_3}, the critical points of $D_{NNN}$ are given by the solutions of
\begin{align}
	\frac{1}{3}\frac{\partial}{\partial\vartheta} D_{NNN} &= -\cos^2\vartheta\sin\vartheta D_{111}+\left(\cos^3\vartheta-2\cos\vartheta \sin^2\vartheta\right) D_{112}\nonumber\\
	&\qquad+\left(2\cos^2\vartheta \sin\vartheta-\sin^3\vartheta\right)D_{122}+\cos\vartheta \sin^2\vartheta D_{222} = 0.
\end{align}
For vanishing $D_{122}$, one solution is given by $\vartheta=\pi/2$, i.e.\ $\omega$ has a critical point at $N^i=n_2^i$. It can be checked by a direct calculation that choosing $n_1^i$ different from Eq.\ \eqref{eq:n1_ex_3} as
\begin{align}
	\frac{1}{\sqrt{C_{23}}}\begin{pmatrix}0\\K_3\\-K_2\end{pmatrix}\quad\text{or}\quad \frac{1}{\sqrt{C_{13}}}\begin{pmatrix}K_3\\0\\-K_1\end{pmatrix}
\end{align}
also results in $D_{122}=0$ and therefore gives the other two critical points of $\omega$. In total, we find
\begin{align}
	\omega(\vartheta_1) &= \frac{-9e^{-2K}}{(K^3)^3d_{333}\left[d_{111}(K^1)^3+d_{222}(K^2)^3\right]} = \frac{9}{32}e^{10K}d_{333}^2\frac{\left(d_{111}d_{222}K^1K^2\right)^4}{C_{12}\det g^3} > 0\displaybreak[0]\\
	\omega(\vartheta_2) &= \frac{-9e^{-2K}}{(K^2)^3d_{222}\left[d_{111}(K^1)^3+d_{333}(K^3)^3\right]}
		= \frac{9}{32}e^{10K}d_{222}^2\frac{\left(d_{111}d_{333}K^1K^3\right)^4}{C_{13}\det g^3} > 0\displaybreak[0]\\
	\omega(\vartheta_3) &= \frac{-9e^{-2K}}{(K^1)^3d_{111}\left[d_{222}(K^2)^3+d_{333}(K^3)^3\right]} = \frac{9}{32}e^{10K}d_{111}^2\frac{\left(d_{222}d_{333}K^2K^3\right)^4}{C_{23}\det g^3} > 0.
\end{align}
At all three critical points $\omega$ is always positive as long as $g>0$ holds.

\subsection{Partially factorizing models for $p=3$}
\label{sec:almost_fact_3}
Another $p=3$-dimensional example which can be treated in more detail is given by a volume factorizing as
\begin{align}
	\mathcal{V} = -\frac{1}{6}d_iK^i d_{jk}K^jK^k,
\end{align}
where $d_i$ is a vector and $d_{jk}$ is a symmetric non-degenerated matrix. $d_{ijk}=d_{(i}d_{jk)}$ has two vanishing tensorial eigenvalues and by Eq.\ \eqref{eq:lambda} $\omega$ therefore has to vanish at two of its critical points:
\begin{align}
	\label{eq:almost_fact_firsttwo}
	\omega(\vartheta_1)=\omega(\vartheta_2)=0.
\end{align}
To make this more explicit, we can choose coordinates such that
\begin{align}
	\label{eq:vector_d}
	d_i = \begin{pmatrix}d_1\\0\\0\end{pmatrix}.
\end{align}
Then the corresponding zero eigenvectors are easily computed to be
\begin{align}
	\label{eq:vplusminus}
	v_\pm = C_\pm\begin{pmatrix}0\\-d_{23}\pm\sqrt{d_{23}^2-d_{22}d_{33}}\\d_{22}\end{pmatrix},
\end{align}
where $C_\pm$ are normalization constants.\par\medskip

We now derive an expression for $\omega(\vartheta_3)$ which depends only on the scalar product of $v_+$ and $v_-$. For this, choose $C_\pm$ such that $v_\pm$ satisfies $g_{ij}v_\pm^iv_\pm^j=1$. Note that $v_\pm$ are unit vectors only if they are real, because their norm is given by $\sqrt{g_{ij}v_\pm^i\overline{v}_\pm^j}$. Then it holds (due to $v_\pm^i$ being zero eigenvectors of $d_{ijk}$) that
\begin{align}
	1 = g_{ij}v_\pm^i v_\pm^j = e^K d_{ijk}v_\pm^iv_\pm^jK^k+(v_\pm^iK_i)^2 = (v_\pm^iK_i)^2.
	\end{align}
This implies (possibly after changing the orientation) that
	\begin{align}
		v_\pm^i = \frac{1}{3}K^i+\sqrt{\frac{2}{3}} u_\pm^i,
	\end{align}
where $u_\pm$ are normalized vectors orthogonal to $K^i$. According to Eq.\ \eqref{eq:v_decomp_p} the vectors $u_\pm$ are the extremizers of $\omega$ corresponding to the critical points in Eq.\ \eqref{eq:almost_fact_firsttwo}. Now we make the ansatz
\begin{align}
	v^i = \alpha K^i +\beta v_+^i+\gamma v_-^i
\end{align}
for the eigenvalue problem Eq.\ \eqref{eq:ev_problem}. Plugging $v^i$ into the eigenvector equations gives
\begin{align}
	&-2\alpha^2 K_i +2\alpha e^K d_{ijk}K^j\left(\beta v_+^k+\gamma v_-^k\right)+2\beta\gamma e^Kd_{ijk} v_+^jv_-^k \displaybreak[0]\nonumber\\
	&\quad\quad=\lambda\left(3\alpha^2 K_i+2\alpha\beta\left(\frac{2}{3}K_i+v_{+i}\right)+2\alpha\gamma\left(\frac{2}{3}K_i+v_{-i}\right)+\beta^2\left(\frac{2}{3}v_{+i}+\frac{1}{3}K_i\right)\right.\nonumber\\
	&\quad\quad\qquad\left.+\gamma^2\left(\frac{2}{3}v_{-i}+\frac{1}{3}K_i\right)+\frac{2}{3}\beta\gamma\left(\eta K_i+v_{+i}+v_{-i}\right)\vphantom{3\alpha^2}\right),
\end{align}
where we defined
\begin{align}
	\eta:=g_{ij}v_+^iv_-^j.
\end{align}
By multiplying with $K^i$, $v_+^i$ and $v_-^i$ we obtain the system of equations
\begin{align}
	&-6\alpha^2-4\alpha\left(\beta+\gamma\right)+2\beta\gamma\left(\eta-1\right)\nonumber\\
	&\quad= \lambda\left(9\alpha^2+6\alpha\left(\beta+\gamma\right)+\frac{5}{3}\left(\beta^2+\gamma^2\right)+\frac{2}{3}\beta\gamma\left(2+3\eta\right)\right)\displaybreak[0]\\
	&-2\alpha^2+2\alpha\gamma\left(\eta-1\right)\nonumber\\
	&\quad= \lambda\left(3\alpha^2+\frac{10}{3}\alpha\beta+\frac{2}{3}\alpha\gamma\left(2+3\eta\right)+\beta^2+\frac{1}{3}\gamma\left(\gamma+2\beta\right)\left(2\eta+1\right)\right)\displaybreak[0]\\
	&-2\alpha^2+2\alpha\beta\left(\eta-1\right)\nonumber\\
	&\quad= \lambda\left(3\alpha^2+\frac{10}{3}\alpha\gamma+\frac{2}{3}\alpha\beta\left(2+3\eta\right)+\gamma^2+\frac{1}{3}\beta\left(\beta+2\gamma\right)\left(2\eta+1\right)\right).
\end{align}
This system has a solution with a (potentially) real $\lambda$, where $\beta=\gamma$ and $\alpha$ and $\lambda$ are given by relatively complicated expressions. Fortunately, the $\omega$ corresponding to this eigenvalue simplifies considerably and is given by
\begin{align}
	\label{eq:omega_eta}
	\omega(\vartheta_3) &= -27\eta\frac{(\eta-1)^2}{(1+3\eta)^3}.
\end{align}
Because $v_\pm^i K_i=1$ it always holds that $\eta>-1/3$ and therefore
\begin{align}
	\omega(\vartheta_3) > 0\quad\Leftrightarrow\quad \eta<0.
\end{align}
We can now compute $\eta=g_{ij}v_+^iv_-^j$ for $v_\pm$ defined in Eq.\ \eqref{eq:vplusminus} and then use Eq.\ \eqref{eq:omega_eta} to obtain
\begin{align}
	\label{eq:omega_two_ev}
	\omega(\vartheta_3) = \frac{1}{2}e^{7K}\left(d_{133}d_{122}-d_{123}^2\right)^2\det d_{ij}\frac{\left(d_1K^1\right)^3}{\det g^3}.
\end{align}
The quantity $4{\left(d_{133}d_{122}-d_{123}^2\right)}^2$ turns out to be the first Aronhold invariant $S=S\left[d_{ijk}\right]$, the degree-4 invariant of cubic polynomials in three dimensions we used at the end of Section \ref{sec:product_formula} to express the discriminant $\Det\left[d_{ijk}\right]$ in terms of simpler invariants. By performing arbitrary rotations to eliminate the restriction in Eq.\ \eqref{eq:vector_d} and using the invariance of $S$, $\det g$ and $\det d_{ij}$ under these rotations, it follows that the general formula has to be
\begin{align}
	\label{eq:omega_two_ev_gen}
	\omega(\vartheta_3) = \frac{1}{8}e^{7K}S\left[d_{ijk}\right]\,\det d_{ij}\frac{\left(d_iK^i\right)^3}{\det g^3}.
\end{align}
Since in the special coordinate system considered above $S$ can be written as the square of a real number, it has to be positive in this class of models. Using
\begin{align}
	e^{-K}=\mathcal{V} = -\frac{1}{6} d_iK^i d_{jk}K^jK^k,
\end{align}
it follows that
\begin{align}
	\label{eq:omega_two_ev_gen_V}
	\omega(\vartheta_3) = -\frac{3}{4}e^{6K}S\left[d_{ijk}\right]\frac{\det d_{ij}}{d_{jk}K^jK^k}\frac{\left(d_iK^i\right)^2}{\det g^3}.
\end{align}
In particular, if $d_{ij}$ is positive or negative definite, the factor $\frac{\det d_{ij}}{d_{jk}K^jK^k}$ is always positive and $\omega$ is negative. If $d_{ij}$ is indefinite, the sign of this factor constitutes a simple and direct constraint on the allowed values of $K^i = -\left(T^i+\overline{T}^i\right)$.

\section{Conclusion}
\label{sec:conclusion}
In this paper we studied constraints on moduli spaces of heterotic string compactifications imposed by the required existence of metastable classical de Sitter vacua, assuming that only moduli fields participate in supersymmetry breaking. We concentrated on three-dimensional moduli spaces and gave the generalization of the two-dimensional result Eq.\ \eqref{eq:omega_twofields}, which has first been derived in \cite{Covi:2008ea}, in Eq.\ \eqref{eq:omega_prod}. This equation encodes a rather non-trivial result: If the sign of a degree-12 invariant -- the discriminant -- of the Calabi-Yau intersection tensor is positive, the metastability condition is automatically satisfied on all physically acceptable points on the moduli space. Numerical studies suggest that if the discriminant is negative, metastable de Sitter vacua still exist in the generic case, but only for a restricted set of moduli configurations. As briefly discussed in section \ref{sec:general_case}, generalizations of Eq.\ \eqref{eq:omega_prod} seem to exist for arbitrary-dimensional moduli spaces, raising the question of the existence of a more intuitive interpretation of the discriminant.\par\medskip

We also studied specific examples of three-dimensional moduli spaces. For moduli spaces with dimension $p>2$, the metastability analysis is difficult to carry out explicitly and the result in general depends on non-topological properties of the Calabi-Yau, in this case its K\"ahler structure (cf.\ Eq.\ \eqref{eq:pertubation_example} and Eq.\ \eqref{eq:omega_two_ev_gen_V} and the numerical result in Fig.\ \ref{fig:heterotic}).\par\medskip

This complication already appears in the three-dimensional case. While the reduction of the problem from Eq.\ \eqref{eq:omega_heter_p_gen} to Eq.\ \eqref{eq:crit_points_3} essentially reduces the problem to the task of finding the roots of a cubic polynomial, the extraction of meaningful results has not yet been successful in the general case. On the other hand, the class of Calabi-Yaus studied in Section \ref{sec:almost_fact_3} constitutes a promising candidate for further studies: It naturally generalizes the class of factorizable models, i.e.\ models with a volume of the form
\begin{align}
  \mV = -\frac{1}{6}K^1 d_{ab}K^aK^b,
\end{align}
where $a,\,b$ run from $2$ to $p$, while not suffering from the fact that factorizable models do not allow metastable de Sitter vacua without invoking higher-order corrections or additional tree-level contributions\cite{Covi:2008ea}.\par\medskip

\section*{Acknowledgements}

I would like to thank Laura Covi for helpful discussions and comments.

\begin{appendix}

\renewcommand{\theequation}{\thesection.\arabic{equation}}
\section{The complex phase}
\label{sect:complex_phase}
In this appendix we demonstrate that for $p=3$-dimensional moduli spaces of heterotic string models the phases $\varphi_1$ and $\varphi_2$ in Eq.\ \eqref{eq:omega_heter_p_gen} can safely be set to zero in the metastability analysis. To see this, we parameterize $N^i$ as in Eq.\ \eqref{eq:param_N_p} and write $\omega$ as
\begin{align}
	\omega &= -\frac{3}{2}+\frac{3}{2}c_1^4 C_{1111}+\frac{3}{2}c_2^4 C_{2222}+6\left(c_1^3c_2C_{1112}+c_1c_2^3C_{1222}\right)\cos(\varphi_1-\varphi_2)\nonumber\\
	&\quad\quad +c_1^2c_2^2\left[C_{1122}+5C_{1212}+(2C_{1122}+C_{1212})\cos(2\varphi_1-2\varphi_2)\right],
\end{align}
where
\begin{align}
	C_{\beta\gamma\delta\eta}:=\sum_{\alpha=1}^2 D_{\alpha\beta\gamma}D_{\alpha\delta\eta}+\frac{1}{3}\delta_{\beta\gamma}\delta_{\delta\eta}.
\end{align}
By swapping the sign of $c_1$, the term $c_1^3c_2C_{1112}+c_1c_2^3C_{1222}$ can always be made a positive contribution to $\omega$, which is maximal if $\varphi_1-\varphi_2=0$. The term proportional to $2C_{1122}+C_{1212}$ can potentially give a negative contribution, which can be reduced (or even turned to a positive one) if the complex phases do not vanish. Hence, we find that the global maximum of $\omega$ in the three-dimensional case $p=3$ can only have a non-vanishing (non-global) phase $\varphi_1-\varphi_2$ if
\begin{align}
	\label{eq:phi_cond_2_gen}
	2C_{1122}+C_{1212} = \frac{2}{3}+D_{112}^2+D_{122}^2+2D_{111}D_{122}+2D_{112}D_{222} < 0.
\end{align}
The quantity $D_{122}$ changes sign if the vectors $n_1^i$ and $n_2^i$ are rotated into each other by $\pi$, implying that we can choose a basis $n_1^i,\,n_2^i$ such that $D_{122}=0$. In this basis, Eq.\ \eqref{eq:phi_cond_2_gen} reads
\begin{align}
  \label{eq:phi_cond_2}
  \frac{2}{3}+D_{112}^2+2D_{112}D_{222} < 0.
\end{align}
This is only fulfilled if
\begin{align}
	-D_{222}-\sqrt{D_{222}^2-\frac{2}{3}}<D_{112}<-D_{222}+\sqrt{D_{222}^2-\frac{2}{3}}.
\end{align}
In particular, it has to hold that
\begin{align}
	D_{222}^2>\frac{2}{3},
\end{align}
implying that $\omega(c_1=0,\,c_2=1,\varphi_\alpha=0)>0$. Thus, if the global maximum of $\omega$ is attained at $\varphi_1-\varphi_2\neq0$, there will always be another critical point of $\omega$ with $\varphi_1-\varphi_2=0$ at which $\omega$ is still positive. Finally, a global phase $\varphi_1=\varphi_2$ always drops out of $\omega$, proving the claim.

\section{Discarding the fourth critical point}
\label{sec:fourth_crit}
After restricting to $p=3$ and setting the complex phases to zero, we found four critical points of $\omega$ as a function of $\vartheta$. We now show that, as has been claimed in Section \ref{sec:3_dim}, that one of these, namely the one given by Eq.\ \eqref{eq:second_poss}, can be discarded a priori in the search for the global maximum of $\omega$. As in the last section, we use the freedom in choosing $n_1^i$ and $n_2^i$ to set $D_{122}=0$. Then by explicitly solving Eq.\ \eqref{eq:second_poss} for the critical point $\vartheta=\vartheta_4$ and plugging the result back into $\omega$ we find
\begin{align}
  \omega(\vartheta_4)=-1+\frac{3}{2}\left[D_{NNN}^2(\vartheta_4)+D_{NNM}^2(\vartheta_4)\right] = -1+\frac{3}{2}\frac{D_{112}^2\left(D_{112}+D_{222}\right)^2+D_{111}^2D_{222}^2}{D_{111}^2+\left(D_{112}+D_{222}\right)^2}.
\end{align}
To prove that this can never constitute the global maximum of $\omega$, we have to show that
\begin{align}
  \max_{\vartheta\in[0, 2\pi]}D_{NNN}^2(\vartheta)\geq\frac{D_{112}^2\left(D_{112}+D_{222}\right)^2+D_{111}^2D_{222}^2}{D_{111}^2+\left(D_{112}+D_{222}\right)^2},
\end{align}
where
\begin{align}
  \label{eq:defA}
  D_{NNN} &= \cos^3\vartheta D_{111}+3\cos^2\vartheta\sin\vartheta D_{112}+\sin^3\vartheta D_{222}.
\end{align}
We abbreviate $x=\tan \vartheta$, $\lambda=D_{111}/D_{112}$ and $\mu=D_{222}/D_{112}$. If $|\mu|\geq1$, setting $\vartheta=\pi/2$ shows the claim. If $|\lambda|\geq1$, we can set $\vartheta=0$ and are done. Let us thus assume that $|\lambda|<1$ and $|\mu|<1$. Then
\begin{align}
  \frac{D_{112}^2\left(D_{112}+D_{222}\right)^2+D_{111}^2D_{222}^2}{D_{111}^2+\left(D_{112}+D_{222}\right)^2} = D_{112}^2+\frac{D_{111}^2\left(D_{222}^2-D_{112}^2\right)}{D_{111}^2+\left(D_{112}+D_{222}\right)^2} < D_{112}^2
\end{align}
and the claim follows if there is an $x$ such that
\begin{align}
	\label{eq:ungl_reduz}
  \frac{1}{\left(1+x^2\right)^3}\left(\lambda+3x+\mu x^3\right)^2 \geq 1.
\end{align}
An extremum in $x$ of the left-hand side of Eq.\ \eqref{eq:ungl_reduz} must satisfy
\begin{align}
	\label{eq:ungl_max}
  \lambda+3x_0+\mu x_0^3 = (1+x_0^2)\left(\frac{1}{x_0}+\mu x_0\right).
\end{align}
This equation is in fact only quadratic in $x_0$ and is solved by
\begin{align}
  \label{eq:ungl_extr}
  x_0 = \frac{-\lambda\pm\sqrt{\lambda^2-4\mu+8}}{2(2-\mu)}.
\end{align}
Taking the `$+$' solution if $\lambda\geq0$ and the `$-$' solution if $\lambda<0$ we see that (by concavity of the square root)
\begin{align}
  x_0^2\leq\frac{1}{2-\mu},
\end{align}
so
\begin{align}
	\label{eq:inequ_fraction}
  \frac{1}{1+x_0^2}\geq \frac{2-\mu}{3-\mu}.
\end{align}
The next step is to show that $\left(\frac{1}{x_0}+\mu x_0\right)$ grows monotonically for $0\leq\lambda\leq 1$. This can be seen from
\begin{align}
	\label{eq:inequ_paranth}
	\frac{\partial}{\partial\lambda}\left(\frac{1}{x_0}+\mu x_0\right) = \left(-\frac{1}{x_0^2}+\mu\right)\frac{\partial}{\partial\lambda}x_0>0,
\end{align}
because
\begin{align}
	\frac{\partial}{\partial\lambda}x_0 = \frac{1}{2(2-\mu)}\left(\frac{\lambda}{\sqrt{\lambda^2+\varepsilon^2}}-1\right)<0
\end{align}
with $\varepsilon^2=8-4\mu>0$ and
\begin{align}
	-\frac{1}{x_0^2}+\mu \leq 2\mu-2 < 0.
\end{align}
This finally gives for $0\leq\lambda\leq 1$
\begin{align}
  \label{eq:ungl_fertig}
  \frac{1}{\left(1+x_0^2\right)^3}\left(\lambda+3x_0+\mu x_0^3\right)^2 &\overset{\eqref{eq:ungl_max}}{=} \frac{1}{1+x_0^2}\left(\frac{1}{x_0}+\mu x_0\right)^2\\
  	&\overset{\eqref{eq:inequ_fraction}}{\geq} \frac{2-\mu}{3-\mu}\left(\frac{1}{x_0}+\mu x_0\right)^2\\
  	&\overset{\eqref{eq:inequ_paranth}}{\geq} \frac{2-\mu}{3-\mu}\left(2-\mu+2\mu+\frac{\mu^2}{2-\mu}\right)\\
  	&\,\,\,= \frac{4}{3-\mu} > 1.
\end{align}
The calculation for $\lambda<0$ is analogous. Alternatively, the claim follows by substituting $x\to-x$ in Eq.\ \eqref{eq:ungl_reduz}.

\end{appendix}


\begin{thebibliography}{99}


 \bibitem{hep-th/0007018}
  J.~Maldacena and C.~Nunez,
  {\it Supergravity description of field theories on curved manifolds and a no go theorem},
  Int.~J.~Mod.~Phys.~A {\bf 16} (2001) 822
  [arXiv:hep-th/0007018].
  
 \bibitem{1003.0029}
  T.~Wrase and M.~Zagermann,
  {\it On Classical de Sitter Vacua in String Theory},
  Fortschr.~Phys. {\bf 58} (2010) 906
  [arXiv:1003.0029].

 \bibitem{0711.2512}
  M.~P.~Hertzberg, S.~Kachru, W.~Taylor and M.~Tegmark,
  {\it Inflationary constraints on type IIA string theory},
  JHEP {\bf 0712} (2007) 095
  [arXiv:0711.2512].

 \bibitem{1107.2925}
  G.~Shiu and Y.~Sumitomo,
  {\it Stability constraints on classical de Sitter vacua},
  JHEP {\bf 1109} (2011) 052
  [arXiv:1107.2925].
  
  \bibitem{1110.0545}
  S.~R.~Green and E.~J.~Martinec and C.~Quigley and S.~Sethi,
  {\it Constraints on String Cosmology},
  Class. Quantum Grav. {\bf 29} (2012)
  [arXiv:1110.0545].

\bibitem{Becker:2002nn}
  K.~Becker, M.~Becker, M.~Haack and J.~Louis,
  {\it Supersymmetry breaking and alpha'-corrections to flux induced potentials},
  JHEP {\bf 0206}, 060 (2002)
  [arXiv:hep-th/0204254].
  
\bibitem{Balasubramanian:2004uy}
  V.~Balasubramanian and P.~Berglund,
  {\it Stringy corrections to Kaehler potentials, SUSY breaking, and the
   cosmological constant problem},
  JHEP {\bf 0411} (2004) 085
  [arXiv:hep-th/0408054].
 
\bibitem{Parameswaran:2006jh}
  S.~L.~Parameswaran and A.~Westphal,
  {\it de Sitter string vacua from perturbative Kaehler corrections and
  consistent D-terms},
  JHEP {\bf 0610} (2006) 079
  [arXiv:hep-th/0602253].

\bibitem{Palti:2008mg}
  E.~Palti, G.~Tasinato and J.~Ward,
  {\it Weakly-coupled IIA Flux Compactifications},
  JHEP {\bf 0806} (2008) 084
  [arXiv:0804.1248].

\bibitem{Berg:2007wt}
  M.~Berg, M.~Haack and E.~Pajer,
  {\it Jumping through loops: on soft terms from large volume compactifications},
  JHEP {\bf 0709} (2007) 031
  [arXiv:0704.0737].

\bibitem{1204.0807}
  F.~F.~Gautason, D.~Junghans and M.~Zagermann,
  {\it On Cosmological Constants from alpha'-Corrections},
  JHEP {\bf 1206} (2012) 029
  [arXiv:1204.0807].

\bibitem{Kachru:2003aw}
 S.~Kachru, R.~Kallosh, A.~Linde and S.~P.~Trivedi,
 {\it De Sitter vacua in string theory},
 Phys.\ Rev.\  D {\bf 68} (2003) 046005
 [arXiv:hep-th/0301240].

\bibitem{Silverstein:2007ac}
  E.~Silverstein,
  {\it Simple de Sitter Solutions},
  Phys.\ Rev.\  D {\bf 77} (2008) 106006
  [arXiv:0712.1196].

\bibitem{0901.0683}
  S.~Krippendorf, F.~Quevedo,
  {\it Metastable SUSY Breaking, de Sitter Moduli Stabilisation and Kähler Moduli Inflation},
  JHEP {\bf 0911} (2009) 039
  [arXiv:0901.0683].
  
\bibitem{1203.1750}
  M.~Cicoli, A.~Maharana, F.~Quevedo and C.~P.~Burgess,
  {\it De Sitter String Vacua from Dilaton-dependent Non-perturbative Effects},
  JHEP {\bf 1206} (2012) 011
  [arXiv:1203.1750].
  
\bibitem{1208.3208}
  J.~Louis, M.~Rummel, R.~Valandro and A.~Westphal,
  {\it Building an explicit de Sitter},
  JHEP {\bf 1210} (2012) 163
  [arXiv:1208.3208].
  
\bibitem{0812.3551}
  C.~Caviezel, P.~Koerber, S.~Kors, D.~Lust, T.~Wrase and M.~Zagermann,
  {\it On the Cosmology of Type IIA Compactifications on SU(3)-structure Manifolds},
  JHEP {\bf 0904} (2009) 010
  [arXiv:0812.3551].

\bibitem{0812.3886}
  R.~Flauger, S.~Paban, D.~Robbins and T.~Wrase,
  {\it On Slow-roll Moduli Inflation in Massive IIA Supergravity with Metric Fluxes},
  Phys.\ Rev.\  D {\bf 79} (2009) 086011
  [arXiv:0812.3886].

\bibitem{Haque:2008jz}
  S.~S.~Haque, G.~Shiu, B.~Underwood and T.~Van~Riet,
  {\it Minimal simple de Sitter solutions},
  Phys.\ Rev.\  D {\bf 79} (2009) 086005
  [arXiv:0810.5328].

\bibitem{0907.2041}
  U.~H.~Danielsson, S.~S.~Haque, G.~Shiu and T.~Van~Riet,
  {\it Towards Classical de Sitter Solutions in String Theory},
  JHEP {\bf 0909} (2009) 114
  [arXiv:0907.2041].

\bibitem{1103.4858}
  U.~H.~Danielsson, S.~S.~Haque, P.~Koerber, G.~Shiu, T.~Van~Riet and T.~Wrase,
  {\it De Sitter hunting in a classical landscape},
  Fortschr.~Phys. {\bf 59} (2011) 897
  [arXiv:1103.4858].

\bibitem{1112.3338}
  X.~Chen, G.~Shiu, Y.~Sumitomo and S.-H.~H.~Tye,
  {\it A Global View on The Search for de-Sitter Vacua in (type IIA) String Theory},
  JHEP {\bf 1204} (2012) 026
  [arXiv:1112.3338].
    
  \bibitem{GRS1}
  M.~Gomez-Reino and C.~A.~Scrucca,
  {\it Locally stable non-supersymmetric Minkowski vacua in supergravity},
  JHEP {\bf 0605} (2006) 015
  [arXiv:hep-th/0602246].
  
  \bibitem{0402088}
 R.~Brustein and S.~P.~de~Alwis,
 {\it Moduli potentials in string compactifications with fluxes: mapping the Discretuum},
 Phys.\ Rev.\  D {\bf 69} (2004) 126006
 [arXiv:hep-th/0402088].
  
    \bibitem{CGGPS}
  L.~Covi, M.~Gomez-Reino, C.~Gross, G.~A.~Palma and C.~A.~Scrucca,
  {\it Constructing de Sitter vacua in no-scale string models without uplifting},
  JHEP {\bf 0903} (2009) 146
  [arXiv:0812.3864].
  
        \bibitem{Covi:2008ea}
  L.~Covi, M.~Gomez-Reino, C.~Gross, J.~Louis, G.~A.~Palma and C.~A.~Scrucca,
  {\it de Sitter vacua in no-scale supergravities and Calabi-Yau string models},
  JHEP {\bf 0806} (2008) 057
  [arXiv:0804.1073].

   \bibitem{Gross-2009-PhdThesis}
  C.~Gross,
  {\it De Sitter Vacua and Inflation in no-scale String Models},
  Dissertation, Universit\"at Hamburg, 2009,
  DESY report: DESY-THESIS-2009-029.
 
  \bibitem{GomezReino:2007qi}
  M.~Gomez-Reino and C.~A.~Scrucca,
  {\it Metastable supergravity vacua with F and D supersymmetry breaking}, 
  JHEP {\bf 0708} (2007) 091
  [arXiv:0706.2785].
  
\bibitem{GomezReino:2008bi}
  M.~Gomez-Reino, J.~Louis and C.~A.~Scrucca,
  {\it No metastable de Sitter vacua in N=2 supergravity with only
  hypermultiplets},
  JHEP {\bf 0902} (2009) 003
  [arXiv:0812.0884].
  
  \bibitem{1985NuPhB.25846C}
  P.~Candelas, G.~T.~Horowitz and A.~Strominger,
  {\it Vacuum configurations for superstrings},
  Nucl.\ Phys.\  B {\bf 258} (1985) 46.

\bibitem{aci}
  L.~Aparicio, D.~G.~Cerde\~no and L.~E.~Ib\'a\~nez,
  {\it Modulus-dominated SUSY-breaking soft terms in F-theory and their test at LHC},
  JHEP {\bf 0807} (2008) 099
  [arXiv:0805.2943].

\bibitem{aciII}
  L.~Aparicio, D.~G.~Cerde\~no and L.~E.~Ib\'a\~nez,
  {\it A 119-125 GeV Higgs from a string derived slice of the CMSSM},
  JHEP {\bf 1204} (2012) 126
  [arXiv:1202.0822].
   
  \bibitem{FSscal}
  D.~Farquet and C.~A.~Scrucca,
  {\it Scalar geometry and masses in Calabi-Yau string models},
  JHEP {\bf 1209} (2012) 025
  [arXiv:1205.5728].
  
  \bibitem{CdlO}
  P.~Candelas and X.~de la Ossa,
  {\it Moduli space of Calabi-Yau manifolds},
  Nucl.\ Phys.\  B {\bf 355} (1991) 455.
  
    \bibitem{Greene199015}
  B.~R.~Greene and M.~R.~Plesser,
  {\it Duality in Calabi-Yau moduli space},
  Nucl.\ Phys.\  B {\bf 338} (1990) 15.
  
      \bibitem{Rathlev:masterthesis}
 D.~Rathlev,
 {\it de Sitter vacua in no-scale supergravity models},
 Master̈́'s thesis, Universit\"at G\"ottingen,
 unpublished.
  
 \bibitem{aronhold}
    S.~Aronhold,
    {\it Zur Theorie der homogenen Funktionen dritten Grades von drei Ver\"anderlichen},
    J.~Reine Angew.~Math. {\bf 39} (1849) 140.

 \bibitem{gelfand1994discriminants}
 I.~M.~Gelfand, M.~M.~Kapranov and A.~V.~Zelevinsky,
  {\it Discriminants, resultants, and multidimensional determinants},
  {Mathematics: Theory \& Applications in Mathematics}, {Birkh{\"a}user}, Boston 1994.
 
 \bibitem{duistermaat2010discrete}
 J.~Duistermaat,
  {\it Discrete Integrable Systems: Qrt Maps and Elliptic Surfaces},
  Springer Monographs in Mathematics, Springer, New York 2010.
  
\end{thebibliography}
\end{document}